\documentclass[showpacs,preprintnumbers,twocolumn,superscriptaddress,aps]{revtex4}
\usepackage{amssymb}
\usepackage[centertags]{amsmath}
\usepackage{txfonts}
\usepackage{epsfig}
\usepackage{bm}
\usepackage{color}
\usepackage{graphicx,graphics}
\usepackage{multirow}

\newcommand{\ovl}[1]{\overline{#1}}

\begin{document}

\title{Hadron production by quark combination in central Pb+Pb collisions at $\sqrt{s_{NN}}= 17.3$ GeV}
\author{Chang-en Shao}
\affiliation{Department of Physics, Qufu Normal University, Shandong 273165,
People's Republic of China}

\author{Jun Song }
\affiliation{Department of Physics, Shandong University, Shandong 250100,
People's Republic of China}

\author{Feng-lan Shao}
\affiliation{Department of Physics, Qufu Normal University, Shandong 273165,
People's Republic of China}

\author{Qu-bing Xie}
\affiliation{Department of Physics, Shandong University, Shandong 250100,
People's Republic of China}

\begin{abstract}
The quark combination mechanism of QGP hadronization is applied to
nucleus-nucleus collisions at top SPS energy. The yields, rapidity
and transverse momentum distributions of identified hadrons in most
central Pb+Pb collisions at $\sqrt{s_{NN}}= 17.3$ GeV are
systematically studied. The calculated results are in agreement with
the experimental data from NA49 Collaboration. The longitudinal and
transverse collective flows and strangeness of the hot and dense
quark matter produced in nucleus-nucleus collisions at top SPS
energy are investigated. It is found that the collective flow of
strange quarks is stronger than light quarks, which is compatible
with that at RHIC energies, and the strangeness is almost the same
as those at $\sqrt{s_{NN}}=$ 62.4, 130, 200 GeV.
\end{abstract}

 \maketitle

\section{Introduction}
Lattice QCD predicts that at extremely high temperature and density,
the confined hadronic matter will undergo a phase transition to a
new state of matter called quark gluon plasma (QGP)
\cite{Susskind1979,qpg2004}. The relativistic heavy ion collisions
can provide the condition to create this deconfined partonic matter
\cite{Gyulassy200530}. In general, two approaches are used to study
the properties of the deconfined hot and dense quark matter produced
in AA collisions. One is studying the high $p_T$ hadrons from
initial hard jets, in which one can recur to the perturbative QCD to
a certain degree \cite{WangXN1998}.  The other is investigating the
properties of thermal hadrons frozen out from the hot and dense
quark matter. For the latter, the hadronization of the hot and dense
quark matter (a typical non-perturbative process) is of great
significance. Only through a reliable hadronization mechanism, can
we reversely obtain various information of QGP properties from the
final state hadrons measured experimentally. The abundant
experimental data \cite{abelev:152301,abelev:2007prl} and
phenomenological studies
\cite{Fries22003prl,Greco2003prl,Fries:2003prc,Greco2003prc,Hwa:2004prc,Hwa:2008prc}
at RHIC energies suggest that quark combination mechanism is one of
the most hopeful candidates.  The two most noticeable results are
the successful explanation of the high baryon/meson ratios and the
constituent quark number scaling of the hadronic elliptic flow in
the intermediate transverse momentum range
\cite{Greco2003prl,Fries:2003prc}, which can not be understood at
all in the partonic fragmentation picture. Recently, the NA49
Collaboration have measured the elliptic flow of identified hadrons
at top SPS energy \cite{Alt2007prc}, and found that the quark number
scaling of elliptic flow was shown to hold also. It immediately
gives us an inspiration of the applicability for the quark
combination at top SPS energy. On the other hand, the NA49
collaboration have found three interesting phenomena around 30A GeV
\cite{Alt08onset}, i.e. the steepening of the energy dependence for
pion multiplicity, a maximum in the energy dependence of strangeness
to pion ratio and a characteristic plateau of the effective
temperature for kaon production. These phenomena are indicative of
the onset of the deconfinement at low SPS energies. One can estimate
via Bjorken method that the primordial spatial energy density in
Pb+Pb collisions at top SPS energy is about 3.0 GeV/$fm^{3}$
\cite{Stock2008}, exceeding the critical energy density (about 1
GeV/$fm^{3}$) predicted by Lattice QCD. Therefore, the deconfined
hot and dense quark matter has been probably created, and we can
extend the quark combination mechanism to SPS energies.

As is well known, hadron yield is one of the most basic and important
observables which can help us to test the understanding of the hadronization
mechanism for the hot and dense quark matter created in the relativistic heavy
ion collisions. In most of recombination/coalescence models, the hadron wave
function is necessary to get the hadron yield. As the wave functions for almost
all hadrons are unknown at present, it is difficult for these models to study
this issue quantitatively \cite{Fries22003prl,Greco2003prl,Molnar03}. In
addition, these models do not satisfy the unitarity which is important to the
issue as well \cite{yang04}. Different from those models, the quark combination
model \cite{Xieqb:1988,Shao2005prc} uses the near-correlation in phase space and
SU$_f$(3) symmetry, instead of hadron wave function, to determine the hadron
multiplicity. In addition, the model satisfies unitarity as well and has
reproduced many experimental data at RHIC
\cite{shao2007prc,Yao:2006fk,WangYF08,Yao08prc}. Therefore, we apply it in this
paper to systematically study the yields, rapidity and transverse momentum
distributions of various hadrons in most central Pb+Pb collisions at
$\sqrt{s_{NN}}= 17.3$ GeV. On one hand, one tests the applicability of the
quark combination mechanism at this collision energy. We note that the first
attempt of the mechanism at SPS energies,  for hadron yields alone, was the
ALCOR model \cite{ALCOR95,Levai2000}. Now, the rich experimental data of hadron
multiplicities and momentum spectra provide an opportunity to make a further,
systematical and even decisive test of the mechanism at SPS energies. On the
other hand, the parton momentum distributions at hadronization, which carry the
information on the evolution of the hot and dense quark mater, are extracted
from the final hadrons at top SPS energy and compared with those at RHIC
energies. We concentrate the comparison on two properties related to strange
hadron production. One is the difference in collective flow between light and
strange quarks, which occurs at RHIC energies \cite{ChenJH2008,WangYF08}. The
other is the strangeness enhancement, a significant property of QGP
\cite{Rafelski1982}.

The paper is arranged as follows.  In the next section, we make a brief
introduction to the quark combination model. In section III, we calculate the
yields and rapidity distributions of identified hadrons in most central Pb+Pb
collisions at $\sqrt{s_{NN}}= 17.3$ GeV. In section IV, the results of
transverse momentum distributions of various hadrons are shown. In section V,
we firstly make a detailed analysis of the longitudinal and transverse
collective flow of the hot and dense quark matter at top SPS energy. Secondly,
the energy dependence of the strangeness in the hot and dense quark matter is
extracted and analyzed. Section VI presents summary.

\section{An introduction of the Quark Combination Model}
The starting point of the model is a color singlet system which consists of
constituent quarks and antiquarks. All kinds of hadronization models demand
that they satisfy rapidity or momentum correlation for quarks in the
neighborhood of phase space. The essence of this correlation is in agreement
with the fundamental requirement of QCD \cite{Xie:ap}. According to QCD, a
$q\overline{q}$ may be in a color octet or a singlet. The color factors
$\langle
(q\bar{q})_8|\frac{-\lambda^{a}\cdot\lambda^{a}}{4}|(q\bar{q})_8\rangle=\frac{1}{6}$
and
$\langle(q\bar{q})_1|\frac{-\lambda^{a}\cdot\lambda^{a}}{4}|(q\bar{q})_1\rangle=-\frac{4}{3}
$, which means a repulsive or an attractive interaction between them. Here
$\lambda^{a}$ are the Gell-Mann matrices. If they are close with each other in
phase space, they can interact with sufficiently time to be in the color
singlet and form a meson. Similarly, a $qq$ can be in a sextet or an
anti-triplet, and the color factors
$\langle(qq)_6|\frac{\lambda^{a}\cdot\lambda^{a}}{4}|(qq)_6\rangle=\frac{1}{3}
$ and $\langle
(qq)_{\bar{3}}|\frac{\lambda^{a}\cdot\lambda^{a}}{4}|(qq)_{\bar{3}}\rangle=-\frac{2}{3}$.
If its nearest neighbor in phase space is a $q$, they form a baryon. If the
neighbor is a $\overline{q}$, because the attraction strength of the singlet is
two times that of the anti-triplet, $q\overline{q}$ will win the competition to
form a meson and leave a $q$ alone to combine with other quarks or antiquarks.
Based on the above QCD and near-correlation in phase space requirements, we had
proposed a quark combination rule(QCR) \cite{Xieqb:1988,Xie:ap} which combines
all these quarks and antiquarks into initial hadrons. When the transverse
momentum of quarks are negligible, all $q$ and $\overline{q}$ can always line
up stochastically in rapidity. The QCR reads as follows:
\begin{enumerate}
\item Starting from the first parton ($q$ or $\overline{q}$) in the line;
\item If the baryon number of the second in the line is of the different type
from the first, i.e. the first two partons are either $q\overline{q}$ or
$\overline{q}q$, they combine into a meson and are removed from the line, go to
point 1; Otherwise they are either $qq$ or $\overline{q}\,\overline{q}$, go to
the next point;
\item Look at the third, if it is of the different type from the first,
the first and third partons form a meson and are removed from the line, go to
point 1; Otherwise the first three partons combine into a baryon or an
anti-baryon and are removed from the line, go to point 1.
\end{enumerate}

The following example shows how the above QCR works
\begin{eqnarray}
\label{example} &&q_1\ovl{q}_2\ovl{q}_3\ovl{q}_4\ovl{q}_5q_6
\ovl{q}_7q_8q_9q_{10}\ovl{q}_{11}q_{12}q_{13}q_{14}\ovl{q}_{15}
q_{16}q_{17}\ovl{q}_{18}\ovl{q}_{19}\ovl{q}_{20}\nonumber\\
&&\rightarrow M(q_1\ovl{q}_2)\;\ovl{B}(\ovl{q}_3\ovl{q}_4\ovl{q}_5)\;
M(q_6\ovl{q}_7)\;B(q_8q_9q_{10})\;M(\ovl{q}_{11}q_{12}) \;\nonumber\\
&&M(q_{13}\ovl{q}_{15})\;B(q_{14}q_{16}q_{17})\;
\ovl{B}(\ovl{q}_{18}\ovl{q}_{19}\ovl{q}_{20}) \nonumber
\end{eqnarray}

If the quarks and antiquarks are stochastically arranged in rapidity, the
probability distribution for $N$ pairs of quarks and antiquarks to combine into
$M$ mesons, $B$ baryons and $B$ anti-baryons is
\begin{equation}
\label{eq1}
X_{MB}(N)=\frac{2N(N!)^2(M+2B-1)!}{(2N)!M!(B!)^2}3^{M-1}\delta_{N,M+3B}.
\end{equation}
Hadronization is the soft process of the strong interaction and is independent
of flavor, so the net flavor number remains constant during the process. In the
quark combination scheme, this means that the quark number for each certain
flavor prior to hadronization equals to that of all initially produced hadrons
after it. Obviously the quark number conservation is automatically satisfied in
the model. It is different from the non-linear algebraic method in ALCOR model
\cite{ALCOR95} where normalization factor for each quark flavor is introduced
with the constraint of the quark number conservation.

The average number of initially produced mesons $M(N)$ and baryons $B(N)$ are
given by
\begin{eqnarray}
\label{eq2}
\langle M(N) \rangle &=&\sum_M\sum_B M X_{MB}(N)\;,\\
\label{eq3} \langle B(N) \rangle &=&\sum_M\sum_B B X_{MB}(N)\;.
\end{eqnarray}
Then the multiplicity of various initial hadrons is obtained according to their
production weights
\begin{equation}
\label{eq4}
 \langle M^{initial}_{j} \rangle=C_{M_{j}}\langle M(N) \rangle,
\hspace*{0.8cm} \langle B^{initial}_{j} \rangle=C_{B_{j}}\langle B(N) \rangle,
\end{equation}
where $C_{M_{j}}$ and $C_{B_{j}}$ are normalized production weights for the
meson $M_{j}$ and baryon $B_{j}$, respectively. If three quark
flavors are considered only, we can obtain the production weights using the SU$_f$(3) symmetry with a
strangeness suppression factor $\lambda_{s}$ \cite{Xieqb:1988,Shao2005prc},
which are listed in Table \ref{weight}. The extension of the symmetry to
excited states, exotic states and more quark flavors is also straightforward
\cite{excit95,Shao2005prc,Yao08prc}.

Considering the decay contributions from the resonances, we can obtain the
yields of final state hadrons
\begin{equation}
\langle{h_{i}^{final}}\rangle=\langle{h_{i}^{initial}}\rangle +\sum\limits_{j}
B_{r}(j\rightarrow i)\langle{h_{j}}\rangle,
\end{equation}
where the $B_{r}(j\rightarrow i)$ is the weighted decay branching ratio for
$h_j$ to $h_i$ \cite{pdg08p355}.

In principle, the hadron production probability should be calculated from the
matrix element $\langle{q}\overline{q}|M\rangle$ for meson or
$\langle{qqq}|B\rangle$ for baryon. However, the wave functions for almost all
hadrons which are governed by the non-perturbative QCD are unknown at present.
It is difficult to study the production of hadrons quantitatively through their
wave functions. In view of this, the hadron production probability in our model
is determined by the SU$_f$(3) symmetry with a strangeness suppression. This
symmetry has been supported by many experiments, particularly by the
coincidence of the observed $\lambda_{s}$ obtained from various mesons and
baryons \cite{Hofmann:1988gy}. Therefore, the model can quantitatively describe
many global properties for the bulk system by virtue of the Monte Carlo method
\cite{Shao2005prc,shao2007prc,Yao:2006fk,WangYF08,Yao08prc,excit95}.

\begin{table}[!tp]
\renewcommand{\arraystretch}{1.5}
\center \caption{The normalized production weight for baryons and mesons in the
$\texttt{SU}_f(3)$ ground state.  ${r_i}$ is the number of strange quarks in
hadron. The ratio of the vector ($J^{P}=1^-$) to pseudoscalar ($J^{P}=0^-$)
meson follows the spin counting, while that of the decouplet
($J^{P}=\frac{3}{2}^+$) to octet ($J^{P}=\frac{1}{2}^+$) baryon suffers a spin
suppression effect; see Ref. \cite{excit95,Shao2005prc} for details. }
\begin{tabular}{l|l}  \hline\hline
\multirow{2}{0.8cm}{$C_M$}&$C_{M_{i}}=\frac{2J_{i}+1}{4(2+\lambda_{s})^2}\lambda_{s}^{r_i}$, except \\
& $C_{\eta}=\frac{2J_{\eta}+1}{4(2+\lambda_{s})^2}\frac{1+2\lambda_{s}^2}{3}$
\hspace*{0.5cm}
 $C_{\eta'}=\frac{2J_{\eta'}+1}{4(2+\lambda_{s})^2}\frac{2+\lambda_{s}^2}{3}$\\
 \hline
\multirow{2}{0.8cm}&
$C_{B_i}=\frac{4}{(2+\lambda_{s})^3(2J_{i}+1)}\lambda_{s}^{r_i}$, except
\\
{$C_B$}&  $C_{\Lambda}=C_{\Sigma^{0}}=C_{\Sigma^{\ast0}}=C_{\Lambda(1520)}=\frac{3}{2\,(2+\lambda_{s})^3}\lambda_{s}$ \\
 \hline\hline
\end{tabular}\label{weight}
\end{table}

When applying the model to describe the hadronizaton of the hot and dense quark
matter produced in heavy ion collisions, the net-baryon quantum number of the
system perplexes the analysis formula of Eq. \ref{eq1} but it can be easily
evaluated in Monte Carlo program. On the other hand, the the transverse
momentum of quarks is not negligible due to the strong collective flow of quark
matter. In principle, we should define the QCR in three-dimensional phase
space, but it is quite complicated to have it because one does not have an
order or one has to define an order in a sophisticated way so that all quarks
can combine into hadrons in a particular sequence. In practice, the combination
is still put in rapidity and meanwhile the maximum transverse momentum
difference $\Delta_p$ between (anti)quarks are constrained as they combine into
hadrons. The transverse spectra of hadrons have a relationship with the quark
spectra as follows (e.g. for meson)
\begin{eqnarray}
\frac{dN_{M}}{d^{2}\mathrm{\textbf{p}_{T}}}\varpropto && \hspace*{-0.5cm} \int
d^{2}\mathrm{\textbf{p}_{1,T}}d^{2}\mathrm{\textbf{p}_{2,T}}
f_{q}(\mathrm{\textbf{p}_{1,T}})f_{\overline{q}}(\mathrm{\textbf{p}_{2,T}})\delta^{2}
(\mathrm{\textbf{p}_{T}}-\mathrm{\textbf{p}_{1,T}}-\mathrm{\textbf{p}_{2,T}})
\nonumber \\
 && \times \, \Theta(\Delta_p-|\mathrm{\textbf{p}^{\ast}_{1,T}}-\mathrm{\textbf{p}^{\ast}_{2,T}}|),
\end{eqnarray}
where $f_{q/\overline{q}}(\mathrm{\textbf{p}_{T}})$  is the transverse momentum
distribution of the quark/antiquark, assumed to be rapidity-independent in
present work. The superscript asterisk denotes the quark momentum in the
center-of-mass frame of formed hadron.  The limitation $\Delta_p$ is treated as
parameter in our study, and fixed to be $\Delta_p=0.3$ GeV for mesons and
$\Delta_p=0.6$ GeV  for baryons both at RHIC and SPS energies. Note that the
spectrum normalization is determined by the multiplicity in Eq. \ref{eq4}, i.e.
the constraint of the parameter $\Delta_p$ on the hadron yield is neglected.

One issue that is often questioned is the energy and entropy conservation in
quark combination process. As the non-perturbative QCD is unsolved, there is no
rigorous theory which can incorporate the partonic phase as well as hadronic
phase, thus it is difficult to justify or condemn this issue in essence at the
moment. As we know, a lot of the experimental phenomena in intermediate
transverse momentum range at RHIC can be explained beautifully only in the
quark combination scenario. It suggests that maybe this 'puzzling' issue does
not exist. As far as the quark combination itself is concerned, there is no
difference for the combination occurred in the different (intermediate or low)
transverse momentum range. Therefore, whether the properties of low $p_T$ hadrons
can be reproduced or not is also a significant test of the quark combination
mechanism, as the vast majority of hadrons observed experimentally are just
these with low transverse momentum.

\section{Hadron yields and rapidity distributions}
In high energy nucleus-nucleus collisions, the energy deposited in the
collision region excites large numbers of newborn quarks and antiquarks from
the vacuum. Subsequently, the hot and dense quark matter mainly composed of
these newborn quarks will expend hydrodynamically until hadronization. The
net-quarks from the colliding nuclei still carry a fraction of beam energy,
thus their evolution is different from the newborn quarks. One part of
net-quarks are stopped in the hot and dense quark matter, and hadronize
together with it. The other part of net-quarks penetrate the hot quark matter,
and run up to the forward rapidity region. The latter, together with small
amount newborn quarks, form the leading fireball. Their hadronization should be
earlier than that of the hot and dense quark matter with a prolonged expansion
stage, and the hadronization outcomes consist of nucleons and small mount of
mesons.

\begin{figure}[!b]
  \includegraphics[width=7cm]{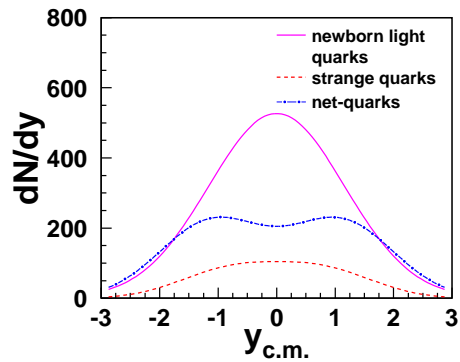}\\
  \caption{(Color online) Rapidity spectra of newborn quarks and net-quarks at hadronization
  in most central Pb+Pb collisions at $\sqrt{s_{NN}}= 17.3$ GeV.
}\label{qdf}
\end{figure}
The current version of quark combination model simulates only the hadronization
of the hot and dense quark matter and subsequently decays of resonances. One
indispensable input is the momentum distributions of thermal quarks and
antiquarks at hadronization, which are the results of the hydrodynamic
evolution in partonic phase. In order to focus attentions on the test of quark
combination mechanism in this and next sections, we reversely extract the quark
distributions by fitting the experimental data in the model. A detailed
analysis of quark distributions at hadronization will be in section V. The
solid and dashed lines in Fig. \ref{qdf} show the rapidity distributions of
newborn light and strange quarks at hadronization respectively, obtained from
the $\pi^{-}$ and $K^{+}$ data \cite{Afanasiev2002prc}.The dotted-dashed line
is the rapidity distribution of net-quarks in the hot and dense quark matter,
which is extracted from net-proton data \cite{Appelshauser1999prl}.

Firstly, we calculate the yields and rapidity densities at midrapidity of
various hadrons in most central Pb+Pb collisions at $\sqrt{s_{NN}}= 17.3$ GeV.
The results are shown in Table \ref{tab1}. From the energy dependence of the
rapidity density for net-baryon \cite{bearden04stop}, one can see that the
nucleus-nucleus collisions at SPS energies exhibit a strong stopping power.
Therefore, the leading particles contribute little to yields and rapidity
distributions of various hadrons.

The calculated yields and rapidity densities of vector meson $\phi$ are shown
to be about twice as high as the experimental data. The results of other
hadrons are basically in agreement with the experimental data, but slight
deviations exist also. The overpredictions of $\phi$ meson may be associated
with the exotic particle $f_{0}(980)$, which has a possible tetraquark
structure containing a strange quark and a strange antiquark \cite{Hirar07prc}.
As a bond state containing strange components, it has a slightly lower mass
than $\phi$ meson but is not included in the SU$_f$(3) ground states. In the
present work, we consider only the production of $36-plets$ of meson and
$56-plets$ of baryon in the SU$_f$(3) ground states, and the excited states and
exotic states are not taken into account. The $f_{0}(980)$ multiplicity is
found to be nearly the same as $\phi$ meson in the $e^{+}e^{-}$ annihilations
\cite{pdg08p355}. The $m_T$ distribution of $f_{0}(980)$ measured by STAR
Collaboration in Au+Au collisions at $\sqrt{s_{NN}}= 200$ GeV is also shown to
be comparable to that of $\phi$ \cite{Fachini2003f980,Adams2005phi}. Therefore,
the overprediction of $\phi$ meson can be removed by incorporating the
$f_{0}(980)$ production.

\begin{table}
\caption{The yields (left) and rapidity densities at midrapidity (right) of
identified hadrons in central Pb+Pb collisions at $\sqrt{s_{NN}}= 17.3$ GeV.
The experimental data are taken from Ref.
\cite{Afanasiev2002prc,Antinorik2005,Alt2008phi,Alt2006prc,Alt08Xi,Mischke03Lam}}
\begin{tabular}{ccc|cc}
\hline \hline
&\multicolumn{2}{c|}{yield}&\multicolumn{2}{c}{$\frac{dN}{dy}|_{y=0}$}
\\ \hline
&data&model&data&model \\
\hline
$\pi^{+}$        & $619\pm17\pm31$  & $566$  & $170.1\pm0.7\pm9$     & 168.2     \\
$\pi^{-}$        & $639\pm17\pm31$  & $630$  & $175.4\pm0.7\pm9$     & 183.5     \\
$K^{+}$          & $103\pm5\pm5$    & $92.5$   & $29.5\pm0.3\pm1.5$    & 27.3      \\
$K^{-}$          & $51.9\pm1.6\pm3$ & $45.3$   & $16.8\pm0.2\pm0.8$     & 15.7     \\
$K^{0}_{s}$      & $75\pm 4$        & 66.7  & $26.0\pm1.7\pm2.6$     & 20.7     \\
$\phi$           & $8.46\pm0.38\pm0.33$& $15.2$     & $2.44\pm0.1\pm0.08$    & 5.26     \\
$p$              &                  & 120  & $29.6\pm0.9\pm2.96$    & 25.9     \\
$\overline{p}$   &                   & 3.2   &$1.66\pm0.17\pm0.17$   & 1.53     \\
$\Lambda$        &  $44.9\pm0.6\pm8$  & 52.9  & $9.5\pm0.1\pm1.0$    & 13.3     \\
$\overline{\Lambda}$
                 &$3.07\pm0.06\pm0.31$ & 2.88  & $1.24\pm0.03\pm0.13$   & 1.35   \\
$\mathrm{\Xi^{-}}$
                 &$4.04\pm0.16\pm0.57$    &4.9   & $1.44\pm0.1\pm0.15$ & 1.43   \\
$\mathrm{\overline{\Xi}^{\,_+}}$
                 &$0.66\pm0.04\pm0.08$   & 0.58    & $0.31\pm0.03\pm0.03$  & 0.26 \\
\hline\hline
\end{tabular}\label{tab1}
\end{table}

\begin{figure*}[!htp]
  \includegraphics[width=17cm]{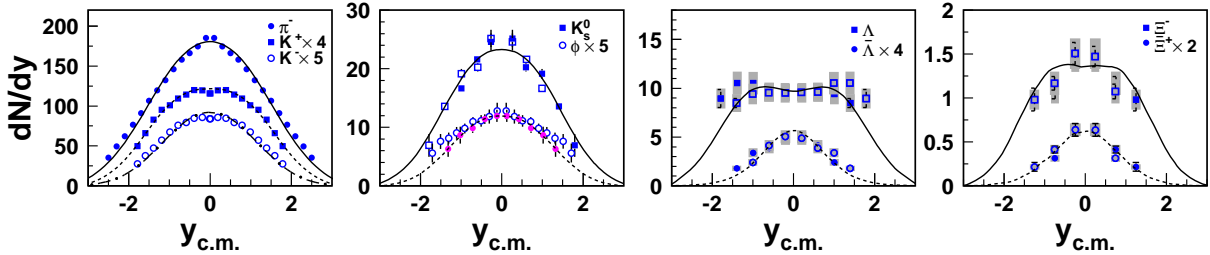}\\
  \caption{(Color online) The scaled rapidity distributions of identified hadrons in most central Pb+Pb
collisions at $\sqrt{s_{NN}}= 17.3$ GeV. The contributions from leading
particles are not included. The open circles of $\phi$ data in the second panel
are the latest results measured by NA49 Collaboration \cite{Alt2008phi}, and
filled circles are the previous ones \cite{Afanasiev2000plb}. Other
experimental data are taken from Ref.
\cite{Afanasiev2002prc,Mischke03Lam,Alt08Xi}.  The open symbols of $K_{s}^{0}$,
$\Lambda$ and $\Xi$ show data points reflected around
midrapidity.}\label{had_y}
\end{figure*}

Subsequently, we will calculate the longitudinal rapidity distributions of
various hadrons. Due to the deviations in hadron yields, it is difficult to
directly compare the calculated hadron spectra with the experimental data. In
order to focus attentions on the property of hadron momentum spectra, we will
scale the calculated rapidity densities to the center value of the experimental
data when we show the hadronic rapidity and $p_{T}$ spectra in Fig. \ref{had_y}
and \ref{hpt173} respectively, thereby removing these deviations in hadron
yields.

Pion is the lightest and most abundant hadron produced in AA collisions, and
its momentum distribution can best reflect the global evolution property of the
hot and dense quark matter. In various models of high energy heavy ion
collisions, the reproduction of pion meson is always taken as a paramount test
of models. In Landau hydrodynamic model \cite{Landau}, the rapidity
distribution of pion can be well described and the sound velocity (which is an
important physical quantity standing for the property of the hot and dense
quark matter) can be extracted from the pion distribution. For other hadrons,
such as kaons, protons, $\Lambda$, $\Xi$ and so on, the Landau model can not
describe their rapidity distributions with the same sound velocity or
freeze-out temperature \cite{Mohanty2003prc,Satarov2007prc,Sarkisyan:2006}. For
a systematic description of the rapidity distributions of various hadrons, the
detailed longitudinal dynamics, e.g. the evolution of net-baryon density which
will result in the yield and spectrum asymmetry between hadron and antihadron,
should be included. In addition, the hadronization mechanism is especially
important to describe the differences in the yield and momentum distribution of
various hadron species. Using the extracted quark distributions in Fig.
\ref{qdf}, we calculate the rapidity distributions of pions, kaons,
$\Lambda(\overline{\Lambda})$, $\Xi^{-}(\overline{\Xi}^{_+})$ and $\phi$ in
most central Pb+Pb collisions at $\sqrt{s_{NN}}= 17.3$ GeV. The results are
shown in Fig. \ref{had_y} and are compared with the experimental data. The
calculated rapidity spectrum of $\phi$ meson is narrow than the latest data of
NA49 Collaboration (open circles in the second panel), but is in good agreement
with previous data (filled circles). The rapidity spectra of other hadrons are
well reproduced. One can see that the quark combination mechanism is applicable
for describing the longitudinal distributions of various hadrons at top SPS
energy.

\section{Hadron transverse momentum distributions}
\begin{figure}[!htp]
  \includegraphics[width=7cm]{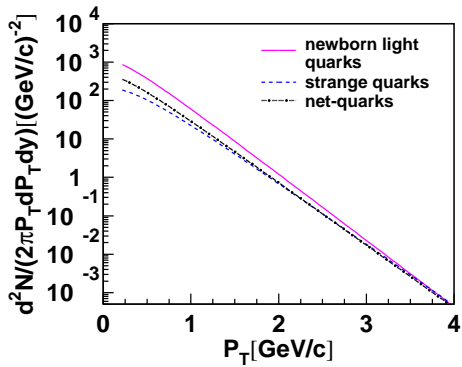}\\
  \caption{(Color online) The transverse momentum distributions of constituent quarks in the midrapidity region at
  hadronization in most central Pb+Pb collisions at $\sqrt{s_{NN}}= 17.3$ GeV. }\label{qpt}
\end{figure}

In this section, we calculate the transverse momentum distributions of various
hadrons in the midrapidity range. In this paper, we only consider the
hadronization of the hot and dense quark matter. The transverse momentum
invariant distribution of constituent quarks at hadronization is taken to be an
exponential form $\exp(-m_{T}/T)$, where $T$ is the slope parameter which is
also called effective production temperature. Fig. \ref{qpt} shows the
midrapidity $p_T$ spectra of constituent quarks at hadronization in most
central Pb+Pb collisions at $\sqrt{s_{NN}}= 17.3$ GeV. The spectra of newborn
light and strange quarks are extracted from the data of $\pi^{+}$ and $K^{+}$
respectively \cite{Alt2008hpt}. The net-quark distribution is fixed by the data
of $K^{-}/K^{+}$ ratio as a function of $p_T$ \cite{Alt2008hpt}. A detailed
analysis of the quark $p_{T}$ spectra will be shown in section V.

\begin{figure*}[!htp]
  \includegraphics[width=16cm]{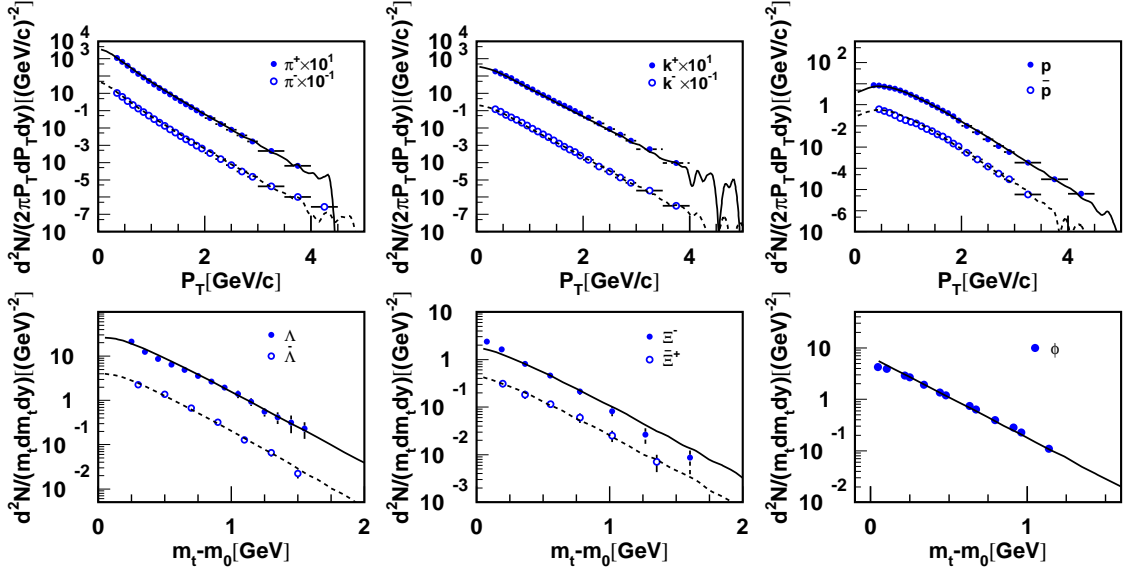}\\
  \caption{(Color online) The scaled transverse momentum distributions of identified hadrons at
  midrapidity in most central Pb+Pb collisions at $\sqrt{s_{NN}}= 17.3$ GeV. Only the combination
  of thermal quarks is taken into account. Solid lines are the calculated results of
  hadrons and dashed lines for antihadrons.  The experimental data are
  from Ref. \cite{Alt2008hpt,Alt08Xi,Alt2008phi}}
  \label{hpt173}
\end{figure*}

Fig. \ref{hpt173} shows the calculated $p_{T}$ spectra of pions, kaons,
protons, $\Lambda(\overline{\Lambda})$, $\Xi^{-}(\overline{\Xi}^{_+})$ and
$\phi$ in most central Pb+Pb collisions at $\sqrt{s_{NN}}= 17.3$ GeV. For
kaons, protons, $\Lambda$ and $\Omega$, the spectral slopes of antihadrons
measured experimentally are all steeper than those of hadrons
\cite{Alt2008hpt,Anticic2004L,Anticic2005O}. However, the spectrum of $\Xi^{-}$
is abnormally steeper than that of $\overline{\Xi}^{_+}$ \cite{Alt08Xi}. Our
predictions are in good agreement with all the data except $\Xi^{-}$. 

The exponential function $\exp(-m_{T}/T)$ is often used experimentally to fit
the transverse momentum distributions of identified hadrons in the low $p_T$
range, and to extract the effective production temperature $T$ of various
hadrons. It is found at top SPS energy that all final-state hadrons except pion
meson have much higher $T$ than the critical temperature \cite{Alessandro2003},
which indicates a strong collective flow at this collision energy. It is
regarded in Ref. \cite{XuNu1998} that this observed flow mainly develops in the
late hadronic rescattering stage. But results in Fig. \ref{had_y} and Fig.
\ref{hpt173} all show that both longitudinal and transverse spectra of various
hadrons can be coherently explained by the same quark distributions,
respectively. It suggests that the observed flow should mainly come from the
expansive evolution stage of the hot and dense quark matter before
hadronization, but not from the post-hadronization stage. In addition, the same
constituent quark spectra contained in light, single- and multi- strange
hadrons also imply that the hot quark matter hadronize into these initial hadrons
almost at the same time, i.e. the hadronization is a rapid process.

\begin{figure*}[!htp]
  \includegraphics[width=16cm]{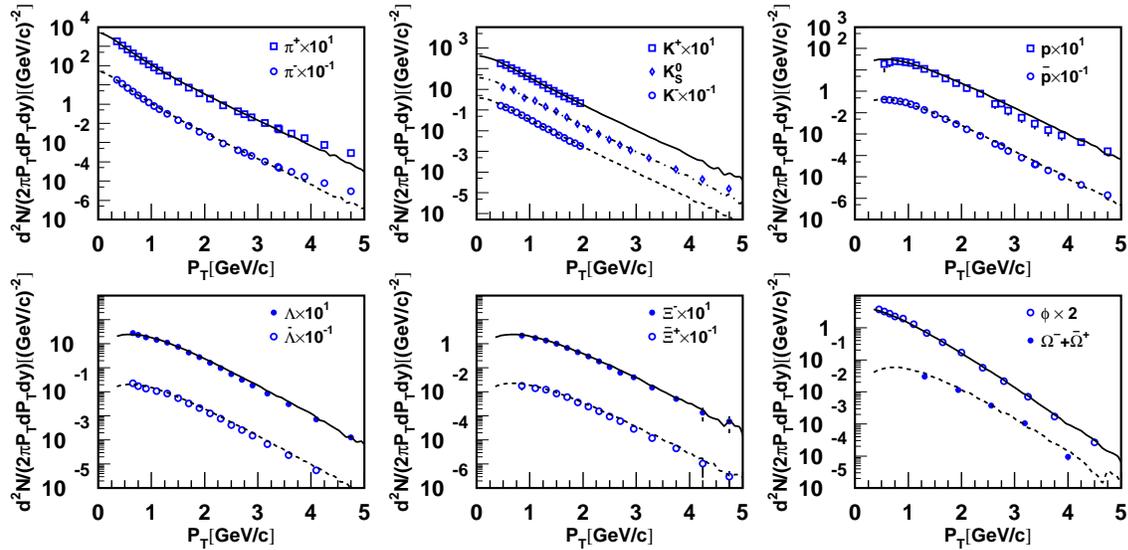}\\
  \caption{(Color online) The transverse momentum distributions of identified hadrons at
  midrapidity in most central Au+Au collisions at $\sqrt{s_{NN}}= 200$ GeV.
  Only the combination of thermal quarks is taken into account. Solid lines
   are the calculated results of hadrons and dashed lines for antihadrons.
  The experimental data are from Ref. \cite{abelev:152301,Adams07hyperon,Abelev07phiv2}}
  \label{hpt200}
\end{figure*}

\begin{figure*}[!htp]
  \includegraphics[width=16cm]{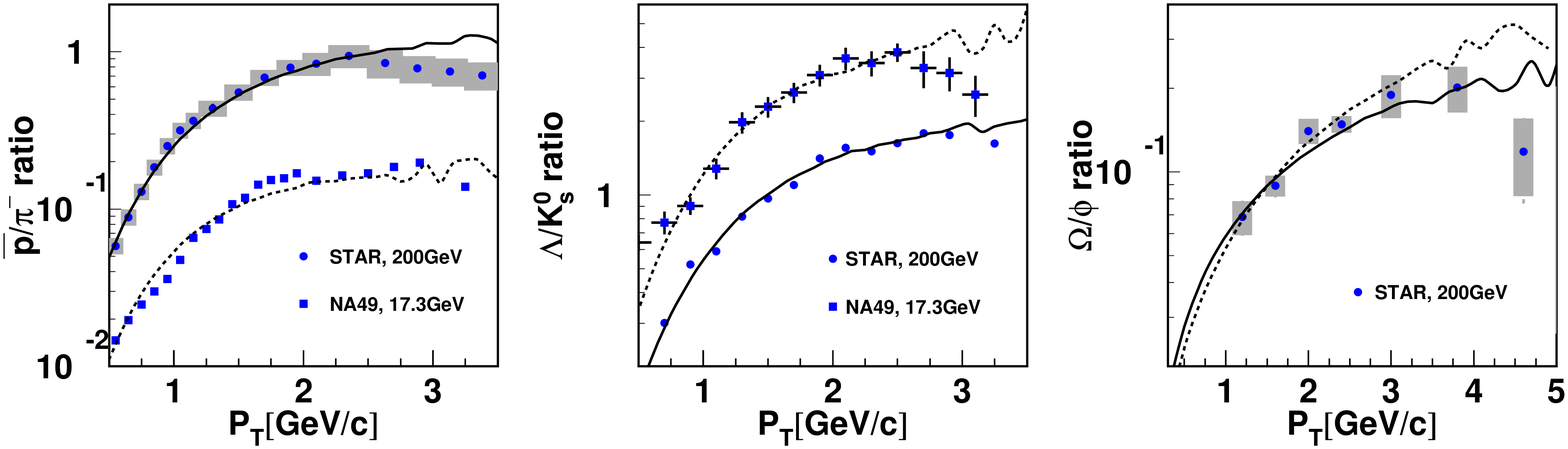}\\
  \caption{(Color online) The ratios of $\overline{p}/\pi^{-}$, $\Lambda/K_{s}^{0}$ and
$\Omega/\phi$ at midrapidity in most central Pb+Pb collisions at
$\sqrt{s_{NN}}= 17.3$ GeV and Au+Au collisions at 200 GeV. Only the combination
  of thermal quarks is taken into account. Solid lines are the calculated results in
  Au+Au collisions and dashed lines for Pb+Pb collisions.
 The experimental data are from Ref.
 \cite{abelev:152301,Abelev06ks0,Abelev07phiv2,Alt2008hpt,Andr06ksL}}\label{bmratio}
\end{figure*}

In Fig. \ref{hpt200}, we calculate the $p_{T}$ spectra of pions, kaons,
protons, $\Lambda(\overline{\Lambda})$, $\Xi^{-}(\overline{\Xi}^{_+})$, $\phi$
and $\Omega(\overline{\Omega})$ in most central Au+Au collisions at top RHIC
energy. The momentum distributions of constituent quarks at hadronization are
taken to be $\exp(-m_{T}/0.375)$ for strange quarks and $\exp(-m_{T}/0.34)$ for
light quarks. The numbers and rapidity spectra of the light and strange quarks
and antiquarks have been obtained in the study of the longitudinal hadron
production \cite{songjun09}. Here, the result of $\phi$ meson is multiplied by
a factor 0.5. One can see that the $p_{T}$ spectra of various hadrons are well
reproduced.

The baryon/meson ratio as a function of $p_{T}$ is sensitive to the
hadronization mechanism. As we know, the observed high baryon/meson ratios in
the intermediate $p_{T}$ range at RHIC energies \cite{abelev:152301} can not be
understood at all in the scheme of parton fragmentation, but can be easily
explained in the quark combination mechanism. Fig. \ref{bmratio} shows the
model predictions of $\overline{p}/\pi^{-}$, $\Lambda/K_{s}^{0}$ and
$\Omega/\phi$ at midrapidity in both central Pb+Pb collisions at
$\sqrt{s_{NN}}= 17.3$ GeV and central Au+Au collisions at$\sqrt{s_{NN}}= 200$
GeV. In the intermediate $p_{T}$ range where the hadron production is dominated
by the combination of thermal quarks, the baryon/meson ratios increase with the
increasing $p_{T}$. One can see that the experimental data in this region are
well reproduced. The falling tendency of measured baryon/meson ratios after
peak position is owing to the abundant participation of jet quarks, which is
beyond the concern of the present paper. Besides the hadronization mechanism,
the baryon/meson ratio in the intermediate $p_{T}$ region is also influenced by
other two factors. One is the nuclear stopping power in collisions. Comparing
with the strong collision transparency at top RHIC energy \cite{bearden04stop},
the strong nuclear stopping at top SPS energy causes the detention of abundant
net-quarks in the midrapidity region. These net-quarks significantly suppress
the production of anti-baryons and enhance that of the baryons. Therefore, the
$\overline{p}/\pi^{-}$ ratio at top SPS energy is much lower than that at top
RHIC energy while the $\Lambda/K_{s}^{0}$ ratio at top SPS energy is higher
than that at top RHIC energy. The other is the momentum distribution of
constituent quarks at hadronization. This can be illustrated by $\Omega/\phi$
ratio because the production of these two hadron species is less influenced by
the net-quarks. The calculated $\Omega/\phi$ ratio shows a weak dependence on
the collision energy in the intermediate $p_T$ range. The well description of
various baryon/meson ratios in such a wide energy range is an indication of the
universality for the quark combination mechanism.

\section{Analysis of parton distributions at hadronization}

The constituent quark distributions at hadronization carry the information on
the evolution of the hot and dense quark matter in partonic phase. In this
section, we focus attentions on the longitudinal and transverse collective
flows and strangeness enhancement of the hot and dense quark matter produced at
top SPS energy.

\subsection{The longitudinal and transverse collective flow}

Due to the thermal pressure, the hot and dense quark matter created in high
energy heavy ion collisions will expand during the evolution before
hadronization. The longitudinal and transverse collective flow of final hadrons
measured experimentally is the exhibition of this early thermal expansion in
the partonic phase. Utilizing the relativistic hydrodynamic evolution of the
hot and dense quark matter, one can obtain the collective flow in quark level
from the extracted quark momentum distributions, and compare it with that at
RHIC energies.

There are two well known hydrodynamic models for the description of the space
time evolution of the hot and dense quark matter produced in heavy ion
collisions. One is Bjorken model \cite{Bjorken84} which supposes that the
collision is transparent, and it is appropriate to extremely high energy
collisions, such as LHC. The other is Landau model \cite{Landau} with an
assumption of the full stopping for nucleus-nucleus collisions. The
longitudinal evolution result is equivalent to the superposition of a set of
thermal sources in rapidity axis, with a (Bjorken) uniform or (Landau) Gaussian
weight. In general, when applying the model to describe the hadron rapidity
distributions, different parameter values are required to make a good fit of
different hadron species \cite{Mohanty2003prc}. In this paper, we apply the
hydrodynamic description to the evolution in quark level, thus the collective
flow of various hadrons can be coherently explained.

One can see from the energy dependence of the net-baryon rapidity distribution
\cite{bearden04stop} that the collisions at top SPS energy are neither full
transparent nor full stopping. The suppositions of nuclear stopping power in
the two models are inappropriate to the nucleus-nucleus collisions at top SPS
energy. For the description of the rapidity distribution for constituent
quarks, one can limit the boost invariance into a finite rapidity range in the
framework of Bjorken model. This modification is often used to analyze the
longitudinal collectivity in hadronic level \cite{Heinz:1993,Mohanty2003prc}.
The rapidity distribution in a isotropic, thermalized fluid element moving with
a rapidity $\eta$ is
\begin{eqnarray}
\frac{dN_{th}}{dy}(y-\eta)&=A\,T_{f}^3 \exp\
\bigg(-\frac{m}{T_{f}}cosh\,(y-\eta)\bigg)\times \nonumber \\
&\bigg (\frac{m^2}{T_{f}^2}+\frac{m}{T_f}
\frac{2}{cosh\,(y-\eta)}+\frac{2}{cosh^2(y-\eta)}\bigg).
\end{eqnarray}
The rapidity distribution of constituent quarks in the hot and dense quark
matter is the longitudinal boost-invariant superposition of multiple isotropic,
thermalized fluid elements
\begin{equation}
\frac{dN}{dy}=\int_{-\eta_{max}}^{\eta_{max}}\frac{dN_{th}}{dy}(y-\eta)\,d\eta,
\end{equation}
 $\eta_{max}$ is the maximal boot rapidity of fluid elements. The average
longitudinal collective velocity is taken to be
$<\beta_{L}>=\tanh(\eta_{max}/2)$.

Here, $T_{f}$ is the temperature of the locally-thermalized hot and dense quark
matter at hadronization. It is taken to be $T_{f}=170$ MeV. $m$ is the
constituent mass of quarks when they evolve to the transition point. It is
taken to be $340$ MeV for light quarks and $500$ MeV for strange quarks. We
have mentioned in above section that the net-quarks, still carrying a fraction
of initial collision energy, have a more complex evolution than hydrodynamic
expansion in longitudinal axis \cite{Wolschin04RDM}. Therefore, we extract the
longitudinal collective flow from the rapidity distribution of newborn quarks.
Since most of the data are measured in the rapidity range about [-1.5, 1.5],
the rapidity spectra of constituent quarks extracted from experimental data are
valid only in this region. Using above equations to fit the rapidity
distribution of newborn constituent quarks in Fig. 1, we obtain
$<\beta_{L}>=0.58$ for light quarks and $<\beta_{L}>=0.65$ for strange quarks.
It is interesting to find that the average longitudinal collective velocity of
strange quarks is obviously greater than that of light quarks.

For the transverse expansion of the hot and dense quark matter, we adopt a
blast-wave model proposed by Heniz \cite{Heinz:1993} within the boost-invariant
scenario. The quarks and antiquarks transversely boost with a flow velocity
$\beta_{r}(r)$ as a function of transverse radial position $r$. $\beta_{r}(r)$
is parameterized by the surface velocity $\beta_{s}$:
$\beta_{r}(r)=\beta_{s}\,\xi^{\,n}$, where $\xi=r/R_{max}$, and $R_{max}$ is
the thermal source maximum radius ($0<\xi<1$). The transverse momentum
distribution of constituent quarks in the hot and dense quark matter can be
equivalently described by a superposition of a set of thermalized fluid
elements, each boosted with transverse rapidity $\rho=tanh^{-1}\beta_{r}$
\begin{equation}
\dfrac{dN}{{2\pi\hspace{1mm}\mathrm{{p}_{T}}
 d{p}_{T}}}=A\int_{0}^{1}\xi\,d\xi\,
 m_{T}\, \times{}
 I_{0}\bigg(\dfrac{{p}_{T}\,sinh\,\rho}{T_{f}}\bigg)
K_{1}\bigg( \dfrac{m_{T}\,cosh\,\rho}{T_{f}}\bigg). \label{thermal-pt}
\end{equation}
Here, $I_{0}$ and $K_{1}$ are the modified Bessel functions.
$m_T=\surd{\overline{{\mathrm{{p}_{T}}}^2+m^2}}$ is the transverse mass of the
constituent quark. The average transverse velocity can be written as
\begin{equation}
\langle \beta_{r}\rangle =\dfrac{\int\beta_{s}\, \xi^{\,n} \xi\, d\xi}{\int \xi
\,d\xi}=\dfrac{2}{n+2}\beta_{s}.
 \label{aver-beta}
\end{equation}
With fixed parameter $n=0.3$, the average transverse velocity $\langle
\beta_r\rangle$ is able to characterize the transverse collective flow of the
hot and dense quark matter. Using the above equations to fit the transverse
momentum distributions of the newborn quarks in Fig. 3, we obtain $\langle
\beta_{r}\rangle=0.41$ for strange quarks and $\langle \beta_{r}\rangle=0.36$
for light quarks. One can see that the $\langle \beta_{r}\rangle$ of strange
quarks is obviously greater than that of light quarks.

Both longitudinal and transverse results at top SPS energy show that
the strange constituent quarks get a stronger collective flow than
the light quarks in the hydrodynamic evolution of partonic matter.
By analyzing the data of multi-strange hadrons \cite{ChenJH2008} and
primary charged hadrons \cite{WangYF08}, the same property is found
also at top RHIC energy. It suggests that the hot and dense quark
matter produced at top SPS energy undergoes a similar hydrodynamic
evolution to that at RHIC energies. It is generally believed that
the decoupled quark and gluon plasma (QGP) has been created at RHIC
energies \cite{Gyulassy05qcdMater}. This similarity of collective
flow property in quark level may be regarded as a signal of QGP
creation at top SPS energy.

\subsection{The enhanced strangeness}
An interesting phenomenon in high energy heavy ion collisions is the enhanced
production of strange hadrons, which is absent in elementary particle
collisions. In relativistic heavy ion collisions, enormous amounts of energy
are deposited in the collision region to create a deconfined hot and dense
quark matter. The multiple scatterings between partons in the hot and dense
quark matter will cause the large production rate of strangeness by
$gg\rightarrow s\bar{s}$ \cite{Rafelski1982}. The high strangeness of the hot
and dense quark matter, after hadronization, finally leads to the abundant
production of the strange hadrons. This phenomenon is regarded as a signal of
QGP creation. As we know, the enhancement of strangeness production at top RHIC
energy is quite obvious \cite{Abelev08enhan}, and it is generally believed that
the QGP has been created at RHIC energies. When the collision energy drops to
the SPS and AGS energies, it is found that the strangeness production peaks at
about 30A GeV and turns into a plateau at higher collision energies
\cite{Alt08onset}. It is an indication of the onset of deconfinement.

\begin{table*}
\caption{The strange suppression factor $\lambda_{s}$ and the calculated hadron
$dN/dy$ at midrapidity in central AA collisions at different energies. The
experimental data are taken from Ref.
\cite{Afanasiev2002prc,Alt2006prc,Alt08Xi,Arsene08Kpi,Abelev62GeV,Taka05sqm,Speltz04sqm,
Adcox130GeV,Adcox02Lam,Adams04Mults,Adler04Light,Adams07hyperon}.}
\begin{tabular}{ccc|cc|cc|cc} \hline \hline
&\multicolumn{2}{c|}{Pb Pb 17.3 GeV }& \multicolumn{2}{c|}{ Au Au 62.4 GeV} & \multicolumn{2}{c|}{Au Au 130 GeV} & \multicolumn{2}{c}{Au Au 200 GeV}\\ \hline
& data&model&data&model &data&model&data&model \\
\hline
$\pi^{+}$   &$170.1\pm0.7\pm9$&  168.3 &$212\pm5.8\pm14$   & $211$  & $276\pm3\pm35.9$    & $268.7$  & $286.4\pm24.2$       & 287.3     \\
$\pi^{-}$   &$175.4\pm0.7\pm9$&  183.5 &$204\pm7.4\pm14$   & $217$  & $270\pm3.5\pm35.1$  & $272.4$  & $281.8\pm22.8$       & 288.3     \\
$K^{+}$     &$29.6\pm0.3\pm1.5$&  27.3 &$33.35\pm2.15$   & $36.3$  & $46.7\pm1.5\pm7$    & $46.6$   & $48.9\pm6.3$         & 48.35      \\
$K^{-}$     &$16.8\pm0.2\pm0.8$&  15.7 &$28.16\pm1.76$   & $29.9$  & $40.5\pm2.3\pm6$     & $43.1$   & $45.7\pm5.2$         & 46.48     \\
$p$         &$29.6\pm0.9\pm2.9$&  25.9 &$27\pm1.8\pm4.6$   & $26.17$  & $19.3\pm0.6\pm3.3$  & $16.45$   & $18.4\pm2.6$         & 17.41     \\
$\overline{p}$
            &$1.66\pm0.17\pm0.16$&1.53 &$11.5\pm1.5\pm2.9$  & $11.15$ & $13.7\pm0.7\pm2.3$  & $11.63$   &$13.5\pm1.8$         & 13.48     \\
$\Lambda$   &$9.5\pm0.1\pm1$&     13.3 &$14.9\pm0.2\pm1.49$   & $13.42$  & $17.3\pm1.8\pm2.7$  & $14.99$  & $16.7\pm0.2\pm1.1$   & 15.76     \\
$\overline{\Lambda}$
            &$1.24\pm0.03\pm0.13$& 1.35&$8.02\pm0.11\pm0.8$   & $6.77$   &$12.7\pm1.8\pm2$   & $11.4$      & $12.7\pm0.2\pm0.9$  & 12.6   \\
$\mathrm{\Xi^{-}}$
            &$1.44\pm0.1\pm0.15$ & 1.43&$1.64\pm0.03\pm0.014$   & $1.63$  &$2.04\pm0.14\pm0.2$  &$1.99$   & $2.17\pm0.06\pm0.19$ & 2.12   \\
$\mathrm{\overline{\Xi}^{\,_+}}$
            &$0.31\pm0.03\pm0.03$& 0.26&$0.989\pm0.057\pm0.057$   & $0.96$   &$1.74\pm0.12\pm0.17$ &$1.67$    & $1.83\pm0.05\pm0.2$  & 1.72 \\
$\mathrm{\Omega+{\overline{\Omega}}}$
            &                    & 0.17&$0.356\pm0.046\pm0.014$   & $0.369$   &$0.56\pm0.11\pm0.05$ &0.551 &$0.53\pm0.04\pm0.04$  &0.539 \\ \hline
$\chi^{2}/ndf$ &\multicolumn{2}{c|}{10.7/7}& \multicolumn{2}{c|}{6.2/8} & \multicolumn{2}{c|}{$1.6/8$} & \multicolumn{2}{c}{$0.88/8$}\\
$\lambda_{s}$ &\multicolumn{2}{c|}{$0.48\pm0.09$} & \multicolumn{2}{c|}{$0.44\pm0.02$}& \multicolumn{2}{c|}{$0.44\pm0.04$} & \multicolumn{2}{c}{$0.42\pm0.025$}\\
 \hline\hline
\end{tabular}\label{lambds}
\end{table*}

In our model, the strangeness of the hot and dense quark matter is
characterized by the suppression factor
$\lambda_{s}=N_{\bar{s}}:N_{\bar{u}}=N_{\bar{s}}:N_{\bar{d}}$, i.e. the ratio
of $s$ quark number to newborn $u$ (or $d$) quark number. By fitting the
experimental data of identified hadrons, we use the model to extract the
$\lambda_{s}$ of hot and dense quark matter at midrapidity in central AA
collisions at four energies $\sqrt{s_{NN}}= 17.3$, 62.4, 130 and 200 GeV, and
the results are shown in Table \ref{lambds}.  The data of midrapidity $dN/dy$
and the calculated results with minimum deviations at different collision
energies are shown also. The statistical uncertainty of $\lambda_{s}$ is fixed
by the twice minimum deviation. 
The model reproduces the hadron yield in reasonably good way, and the
chi-square fit seems to indicate that with increasing collision energy the
agreement with data significantly improves.

One can see that $\lambda_{s}$ in such a broad energy range is nearly unchanged
within statistical uncertainties, exhibiting an obvious saturation phenomenon.
The results of $\lambda_{s}$ are consistent with the grand canonical limit
($\approx 0.45$) of strangeness \cite{Stock2008}.  Using the Bjorken model, one
can estimate that the primordial spatial energy density of the hot and dense
quark matter produced in collisions at top RHIC energy is about $6.0\,
GeV/fm^{3}$ \cite{Stock2008}, double of that in Pb+Pb collisions at top SPS
energy. The difference in primordial energy density is large while the final
strangeness is nearly equal. The hot and dense quark matter created in heavy
ion collisions is shown to be very close to a perfect fluid \cite{visHD08}. It
means that the local relaxation time toward to thermal equilibrium is much
shorter than the macroscopic evolution time of the hot and dense quark matter.
When the hot and dense quark matter evolves to the point of hadronization, the
strangeness abundance should be mainly determined by the current temperature,
irrelevant to the initial energy density and temperature. The same strangeness
is an indication of the universal hadronization temperature for the hot and
dense quark matter with low baryon chemical potential.

\section{Summary}
In this paper, we have systematically studied the longitudinal and transverse
production of various hadrons at top SPS energy in the scheme of quark
combination. Using the quark combination model, we firstly calculate the yields
and rapidity distributions of various hadrons in most central Pb+Pb collisions
at $\sqrt{s_{NN}}= 17.3$ GeV. The calculated results are in agreement with the
experimental data. This indicates that the quark combination mechanism is
applicable in describing the longitudinal hadron production at this collision
energy. Secondly the $p_T$ distributions of various hadrons at top SPS energy
are calculated and compared with the data. It is found that the light, single
and multi-strange hadrons are well reproduced by the same quark distributions.
It indicates that the hadronization of the hot and dense quark matter is a
rapid process. The well reproduced baryon/meson ratios in the intermediate
$p_T$ range at different collision energies are indicative of the universality
for the quark combination mechanism. By fitting the extracted constituent quark
distributions at hadronization with the hydrodynamic scenario, we further
obtain the longitudinal and transverse collective flow of the hot and dense
quark matter produced at top SPS energy. It is found that the strange quarks
get a stronger collective flow than light quarks, which is consistent with that
at RHIC energies.  The strangeness in the hot and dense quark matter produced
at $\sqrt{s_{NN}}= 17.3$, 62.4, 130, 200 GeV are extracted. The almost
unchanged strangeness may be associated with a universal hadronization
temperature for the hot and dense quark matter with low baryon chemical
potential.

\subsection*{ACKNOWLEDGMENTS}
The authors thank Q. Wang, Z. T. Liang and R. Q. Wang for helpful discussions.
The work is supported in part by the National Natural Science Foundation of
China under the grant 10775089 and the science fund of Qufu Normal University.

\end{document}